%% file: main.tex
\documentclass[11pt]{article}
\usepackage[preprint]{acl}
\usepackage{times}
\usepackage{latexsym}
\usepackage{amsmath}
\usepackage{amssymb}
\usepackage[T1]{fontenc}

\usepackage[utf8]{inputenc}

\usepackage{microtype}

\usepackage{inconsolata}

\usepackage{graphicx}
\usepackage{stfloats}
\usepackage{booktabs}
\usepackage{colortbl}
\usepackage{multirow}
\usepackage{tabularx}
\usepackage{listings}
\usepackage{tcolorbox}

\newcommand{\name}{RepoClean}
\newcommand{\eat}[1]{} 

\title{%
  \name{}: Evaluating Coding Agents on\\
  Repository-Level Obsolete Code Cleanup
}

\author{
  Hongyixuan Yuan$^{1}$\thanks{Equal contribution.} \quad
  Qixing Zhou$^{2}$\footnotemark[1] \quad
  Pengcheng Huang$^{3}$\footnotemark[1] \quad
  Qingshuo Ni$^{1}$ \\
  \textbf{Runsong Zhao}$^{3}$ \quad
  \textbf{Shaokai He}$^{1}$ \quad
  \textbf{Xiangpeng Li}$^{1}$\thanks{Corresponding authors.} \quad
  \textbf{Deqiang Ouyang}$^{1}$\footnotemark[2] \\
  $^{1}$Chongqing University \quad
  $^{2}$Nanjing University \quad
  $^{3}$Northeastern University, China \\
  {\small\texttt{hongyixuanyuan@stu.cqu.edu.cn} \quad
  \texttt{qixing.zhou@smail.nju.edu.cn} \quad
  \texttt{pengcheng.neu@outlook.com}} \\
  {\small\texttt{qingshuoni@stu.cqu.edu.cn} \quad
  \texttt{zhaors@mails.neu.edu.cn} \quad
  \texttt{heshaokai@stu.cqu.edu.cn}} \\
  {\small\texttt{xiangpengli.cs@gmail.com} \quad
  \texttt{deqiangouyang@cqu.edu.cn}}
}

\begin{document}

\maketitle

\begin{abstract}
Large language model (LLM) agents have demonstrated strong repository-level programming capabilities. However, as software evolves, repository code may become redundant or obsolete, increasing maintenance costs and potentially introducing functional issues. Agents must therefore identify and remove unnecessary code safely, yet autonomous repository-level code cleanup remains insufficiently evaluated in isolation. To evaluate this capability, we introduce RepoClean, a benchmark of 200 cleanup tasks from 34 popular open-source Python repositories. RepoClean is constructed by automatically reintroducing deleted code from real repository histories into newer versions, preserving current implementations and retaining instances where previously passing tests consequently fail. To restore existing functionality without new feature development, agents must use repository interfaces and cross-file dependencies to determine which historical code to remove and which implementations to preserve. Evaluation of seven models shows that even the best-performing model, GPT-6 Astra, resolves only 23\% of tasks. Further analysis reveals that agents remove substantially less obsolete code when cleanup is not explicitly requested. RepoClean provides a foundation for systematically studying these cleanup challenges and advancing coding agents toward autonomous software maintenance.
\end{abstract}

\section{Introduction}

Coding agents powered by large language models perform repository-level development tasks \citep{NEURIPS2024_5a7c9475,10.1145/3650212.3680384,ICLR2025_a4b6ad6b}. For example, Claude Opus 5.5 \citep{anthropic2026opus55systemcard} achieves reported task success rates of 89.9\% on SWE-Bench Pro \citep{pmlr-v306-deng26f} and 74.2\% on DeepSWE v1.1 \citep{huang2026deepswemeasuringfrontiercoding}. However, software evolution also creates a recurring cleanup problem: repositories may retain dead or redundant historical code, or old logic that conflicts with current implementations, increasing maintenance costs and disrupting functionality. Reliable coding agents therefore need to autonomously determine which historical code to remove. Because current modules may still depend on older implementations, safe cleanup requires tracing these dependencies rather than judging code in isolation.

\begin{figure*}[t]
    \centering
    \includegraphics[width=\textwidth]{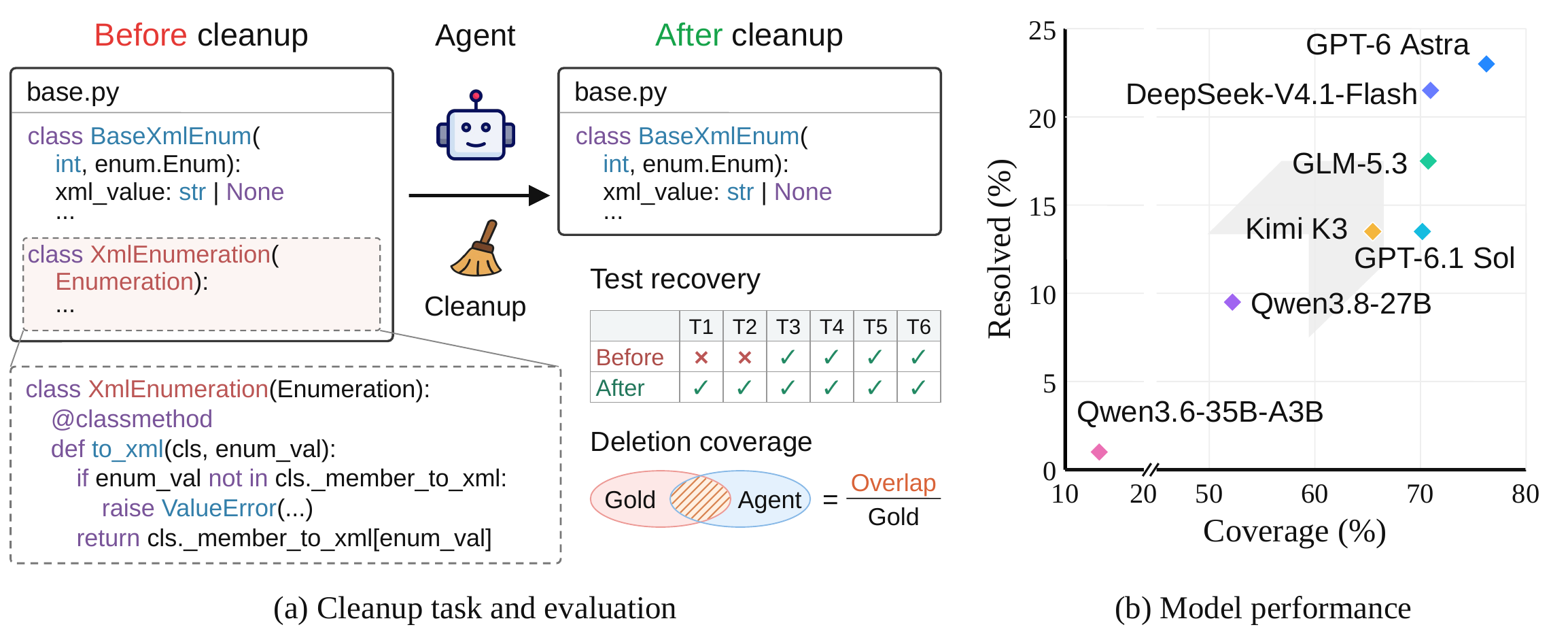}

    \caption{Cleanup task and evaluation in \name{}. (a) An example of obsolete-code removal while preserving the current implementation, with schematic illustrations of test recovery and deletion coverage. (b) Model comparison by task resolution rate and mean deletion coverage on \name{}.}
    \label{fig:repoclean-overview}
\end{figure*}

This raises the question of whether existing coding benchmarks adequately evaluate this cleanup capability. FeatureBench evaluates end-to-end feature development \citep{ICLR2026_25203d1c}, while SWE-EVO and SWE-CHAIN assess repository evolution \citep{le2026sweevobenchmarkingcodingagents,lam2026swechainbenchmarkingcodingagents}. Version updates can require code additions, replacements, and deletions, so overall task success does not isolate cleanup ability. CodeTaste and SmellBench assess the identification and execution of structural improvements \citep{pmlr-v306-thillen26a,lin2026smellbenchfinegrainedevaluationcode}, while CanItDelete evaluates deletion requests within a specified file \citep{ebrahimi2026deletionavoidance}. Historical code cleanup at the repository level remains insufficiently evaluated in isolation. We ask whether agents can use repository interfaces and cross-file dependencies to identify and remove unnecessary historical code while preserving required functionality.

To evaluate this capability, we introduce \name{}, a benchmark of 200 cleanup tasks from 34 popular open-source Python repositories. Table~\ref{tab:benchmark-comparison} compares \name{} with representative coding benchmarks. \name{} is constructed through an automated pipeline that reintroduces previously deleted code from real repository histories into newer repository versions while keeping current implementations intact. The pipeline retains instances where this code causes at least one previously passing validation test file to fail. This automated pipeline supports large-scale construction of executable, verifiable cleanup tasks across repositories. Each task provides a repository, a runtime environment, and cleanup instructions. To restore existing functionality without new feature development, agents must use repository interfaces and cross-file dependencies to determine which historical code to remove and which implementations to preserve. Figure~\ref{fig:repoclean-overview}(a) illustrates an example cleanup task and its evaluation. Functional tests assess whether agents restore affected functionality without introducing regressions. Because passing tests does not establish that obsolete code has been removed, we also measure deletion coverage, the proportion of non-blank source lines deleted by the gold patch that the agent also deletes.

\begin{table*}[t]
\centering
\begingroup
\fontsize{9.5}{11.5}\selectfont
\setlength{\tabcolsep}{2pt}
\renewcommand{\arraystretch}{1.06}
\renewcommand{\tabularxcolumn}[1]{m{#1}}
\begin{tabularx}{\textwidth}{@{}>{\centering\arraybackslash}m{86.4pt}>{\centering\arraybackslash}m{61.1pt}>{\centering\arraybackslash}m{44.5pt}>{\centering\arraybackslash}m{54.5pt}>{\centering\arraybackslash}m{38.1pt}>{\centering\arraybackslash}m{72.4pt}>{\centering\arraybackslash}X@{}}
\toprule
\textbf{Benchmark} & \textbf{Task} & \textbf{Scope} & \textbf{Deletion-only} & \textbf{Instances} & \textbf{Task construction} & \textbf{Evaluation} \\
\midrule
\rowcolor[gray]{0.94}
FeatureBench\newline\citep{ICLR2026_25203d1c} & Feature development & Feature & $\times$ & 200 & Test-guided code extraction & Tests \\[2pt]
Commit0\newline\citep{ICLR2025_1fcefa89} & Library implementation & Repository & $\times$ & 54 & Implementation removal & Tests \\[2pt]
\rowcolor[gray]{0.94}
SWE-EVO\newline\citep{le2026sweevobenchmarkingcodingagents} & Release upgrades & Repository & $\times$ & 48 & Release transition mining & Tests \\[2pt]
SWE-Milestone\newline\citep{pmlr-v306-deng26d} & Continuous evolution & Repository & $\times$ & 98 & Commit grouping & Tests \\
\midrule
\rowcolor[gray]{0.94}
RefactorBench\newline\citep{ICLR2025_6b44ee74} & Refactoring & Repository & $\times$ & 100 & Handcrafted refactorings & AST checks \\[2pt]
CodeTaste\newline\citep{pmlr-v306-thillen26a} & Refactoring & Repository & $\times$ & 100 & Refactoring commit mining & Tests + static checks \\[2pt]
\rowcolor[gray]{0.94}
SmellBench\newline\citep{lin2026smellbenchfinegrainedevaluationcode} & Code smell refactoring & Repository & $\times$ & 294 & Code smell injection & Tests + localization + LLM judge \\[2pt]
CanItDelete\newline\citep{ebrahimi2026deletionavoidance} & Code deletion & File & $\checkmark$ & 200 & Deletion edit mining & Deletion checks \\
\midrule
\rowcolor[gray]{0.94}
\textbf{RepoClean} & \textbf{Historical code cleanup} & \textbf{Repository} & $\boldsymbol{\checkmark}$ & \textbf{200} & \textbf{Historical code reinsertion} & \textbf{Tests + deletion coverage} \\
\bottomrule
\end{tabularx}
\endgroup
\caption{Comparison of RepoClean with representative coding benchmarks. RepoClean evaluates autonomous repository-level cleanup of obsolete historical code. Its automated pipeline reintroduces previously deleted code from real repository histories into newer versions, creating tasks with deletion-only gold patches. Evaluation combines functional tests with deletion coverage.}
\label{tab:benchmark-comparison}
\end{table*}

As shown in Figure~\ref{fig:repoclean-overview}(b), GPT-6 Astra resolves only 23\% of tasks on the full task set, \name{}, with 76.2\% mean deletion coverage. To further examine whether agents autonomously clean up historical code during functional repair, we conduct an instruction ablation on \name{} Lite, a 40-task subset for lower-cost evaluation. Deletion coverage decreases by 28.2--37.1 percentage points across three models when cleanup is not explicitly requested.

Preserving required functionality is another challenge. All seven models cause regressions on \name{}, and 72 of the 115 runs with regressions also make at least one initially failing test file pass. Case analysis further shows that agents can misjudge what to retain or remove, and their self-tests may reinforce rather than expose these errors. These findings highlight the gap between strong coding capabilities and reliable, autonomous code maintenance.

Our contributions are (1) \name{}, a benchmark for autonomous repository-level cleanup of obsolete code while preserving required functionality; (2) an automated pipeline for generating executable cleanup tasks from real repository histories, providing reusable evaluation environments and an evaluation protocol combining functional tests with deletion coverage; and (3) an empirical study across seven models revealing capability gaps in autonomous code cleanup and informing the development of agents for long-term software maintenance.

\section{Related Work}

\paragraph{Repository-Level Coding Benchmarks.}

Commit0, NL2Repo-Bench, and ProgramBench assess codebase construction from specifications or reference behavior \citep{ICLR2025_1fcefa89,pmlr-v306-ding26j,yang2026programbenchlanguagemodelsrebuild}. Beyond construction, SWE-EVO and SWE-CHAIN evaluate software updates across releases \citep{le2026sweevobenchmarkingcodingagents,lam2026swechainbenchmarkingcodingagents}, while SWE-Milestone, SWE-CI, and SlopCodeBench examine successive development tasks through dependent milestones, continuous integration, and changing requirements, respectively \citep{pmlr-v306-deng26d,chen2026swecievaluatingagentcapabilities,orlanski2026slopcodebench}. These benchmarks focus on codebase construction and evolution, rather than historical code cleanup in isolation. \name{} separates obsolete-code cleanup from new feature development using cleanup tasks in repositories where the current version's functionality has already been implemented.

\paragraph{Code Refactoring and Cleanup.}

RefactorBench and SWE-Refactor evaluate agents' execution of specified refactorings \citep{ICLR2025_6b44ee74,xu2026swerefactorrepositorylevelbenchmarkrealworld}, while CodeTaste further examines whether agents can identify the refactorings chosen by human developers within a given scope \citep{pmlr-v306-thillen26a}. SmellBench injects code smells to assess agents' ability to locate and remove structural problems \citep{lin2026smellbenchfinegrainedevaluationcode}. CanItDelete provides a complete target file and a deletion request, assessing deletion completeness and preservation of other code \citep{ebrahimi2026deletionavoidance}. \name{} focuses on repository-level historical code cleanup, evaluating whether agents can use repository context and cross-file dependencies to autonomously identify and remove obsolete historical code. Successful cleanup requires restoring affected functionality without introducing regressions.

\begin{figure*}[t]
    \centering
    \includegraphics[width=\textwidth]{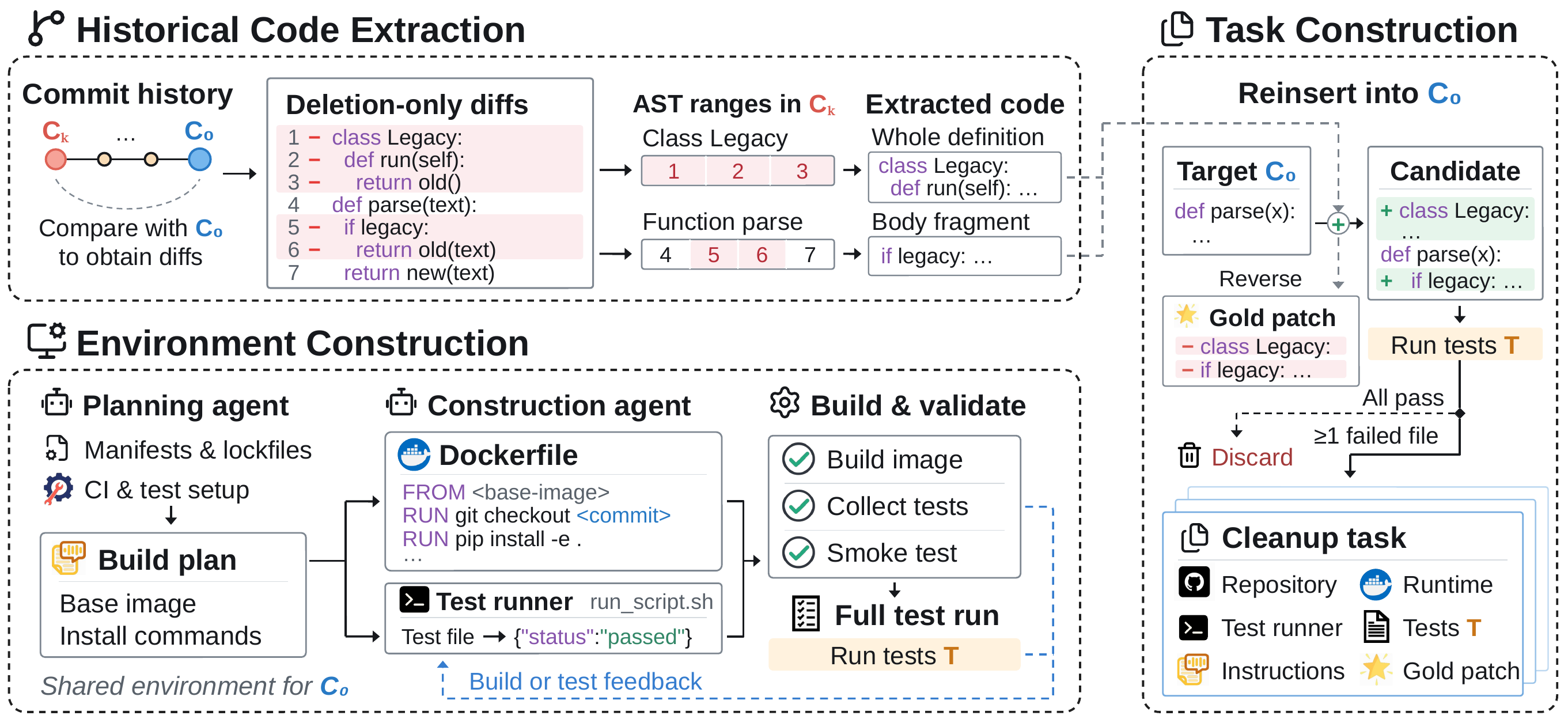}
    \caption{Overview of the \name{} construction pipeline. The pipeline extracts obsolete code from repository histories and uses coding agents to construct executable environments. Reinserting the extracted code into current repository versions creates candidate cleanup tasks. Candidates are retained when the inserted code causes previously passing tests to fail, with its removal providing a verifiable cleanup solution.}
    \label{fig:task-construction}
\end{figure*}

\section{\name{}}

\subsection{Task Definition}

Each task in \name{} provides a repository in which reintroduced historical code disrupts existing functionality, together with a runtime environment and cleanup instructions. The agent must use repository interfaces and cross-file dependencies to autonomously identify and remove redundant code, restore affected functionality, and preserve other functionality. Evaluation checks functional recovery and regressions with validation tests and measures deletion coverage against the gold
patch.

\subsection{Task Construction}
\label{sec:task-construction}

To construct reliable and reproducible tasks for evaluating coding agents, we design a systematic task construction pipeline. Figure~\ref{fig:task-construction} summarizes the pipeline for historical code extraction, executable environment construction, and cleanup task assembly and validation.

\paragraph{Historical Code Extraction.}
We select 100 Python repositories from GitHub that are over one year old, have more than 5,000 stars, and include project configuration or dependency files and a test suite. For each repository, the target version \(C_0\) is the latest default-branch commit at collection time. We sample historical commits \(C_k\) along the default branch's mainline history. For each \(C_k\), we compare Python source files shared by both versions to identify code present in \(C_k\) but no longer present in \(C_0\). We exclude tests, documentation, and configuration files, retaining only diff hunks that contain deletions and no additions.

Abstract syntax tree analysis distinguishes fully deleted functions, methods, or classes from fragments removed within retained functions or methods. Each internal fragment is matched to its corresponding function or method in \(C_0\) using the name and, where applicable, the enclosing class. The fragment is retained only if its insertion position falls within that function or method. Reinsertion only adds historical code, so deleting these additions restores the target version. Appendix~\ref{app:historical-code-extraction} provides detailed rules for sampling historical versions and extracting code.

\paragraph{Environment Construction.}
A planning agent examines repository configuration and CI workflows to select a base image and determine dependency installation and test execution procedures. A construction agent implements this plan as a Docker environment and a repository-specific test script. The Dockerfile specifies how to build the runtime environment, including the base image and dependency installation. Scripts expose a shared interface for discovering and running individual test files, returning structured results even when tests fail. This allows a common evaluator to process results across repositories with different test frameworks. The environment must run tests against the live repository source so that subsequent code edits affect test execution.

Validation first builds the environment and runs a sample of test files. If validation fails, the validation tool returns error information and relevant logs. The construction agent uses the feedback to revise its configuration. The revised Dockerfile or test script is then validated again. Once this check passes, the pipeline runs all collected test files on \(C_0\), each in a fresh container to limit interference between executions. Passing files form the fixed test set \(T\), used to check whether subsequent cleanup introduces functional regressions. Tasks constructed from the same target version share this runtime environment, test script, and test set \(T\). Repositories without a usable environment or passing test files are excluded. Appendix~\ref{app:environment-construction} details the environment construction pipeline.

\paragraph{Task Assembly and Validation.}
The pipeline reinserts the extracted historical code into the target version \(C_0\) using a reinsertion patch, creating the task's initial code state. When different historical versions yield identical reinsertion patches for the same repository and target version, we retain only one task instance.

For the repository copy provided to the agent, we remove the original Git history and initialize a new Git repository with a single commit containing the task's initial code state. To reduce clues from timestamp differences about which files were modified, we set the access and modification times of project files to a common fixed value.

During validation, we first rerun \(T\) on the target version \(C_0\) to confirm that all files in \(T\) pass, then run the same test set on the task's initial code state after reinsertion. If test execution completes and at least one previously passing test file fails, the instance is retained as a cleanup task, with a patch deleting the inserted code serving as the gold patch. Further details on reinsertion, deduplication, workspace preparation, and validation are provided in Appendix~\ref{app:task-assembly-validation}.

\subsection{Evaluation Metrics}
\label{sec:evaluation-metrics}

We evaluate cleanup performance using three metrics: functional recovery score, task resolution, and deletion coverage.

\paragraph{Functional Recovery Score (Recovery).}
For task \(i\), let \(F_i\) be the validation test files that fail before the agent's modifications, \(A_i\) those that pass afterward, and \(\mathrm{P2F}_i\) the number that transition from passing to failing. Recovery is defined as:
\begin{equation}
\mathrm{Recovery}_i =
\begin{cases}
\dfrac{|F_i \cap A_i|}{|F_i|}, & \mathrm{P2F}_i = 0, \\[4pt]
0, & \mathrm{P2F}_i > 0.
\end{cases}
\end{equation}
This metric measures the proportion of initially failing test files that pass after cleanup; the score is zero if any initially passing test file fails.

\paragraph{Task Resolution (Resolved).}
A task receives a Resolved score of one if all validation test files pass after cleanup, and zero otherwise.

\paragraph{Deletion Coverage (Coverage).}
For task \(i\), let \(G_i\) and \(D_i\) denote the sets of non-blank source lines deleted from the original task repository by the gold patch and the agent's patch, respectively. Coverage measures the fraction of gold deletions also made by the agent:
\begin{equation}
\mathrm{Coverage}_i =
\frac{|G_i \cap D_i|}{|G_i|}.
\end{equation}
Figure~\ref{fig:repoclean-overview}(a) illustrates this calculation, with the overlap representing source lines deleted by both patches.
We report the arithmetic mean of per-task scores for all three metrics.

\begin{figure}[t]
    \centering
    \includegraphics[width=\columnwidth]{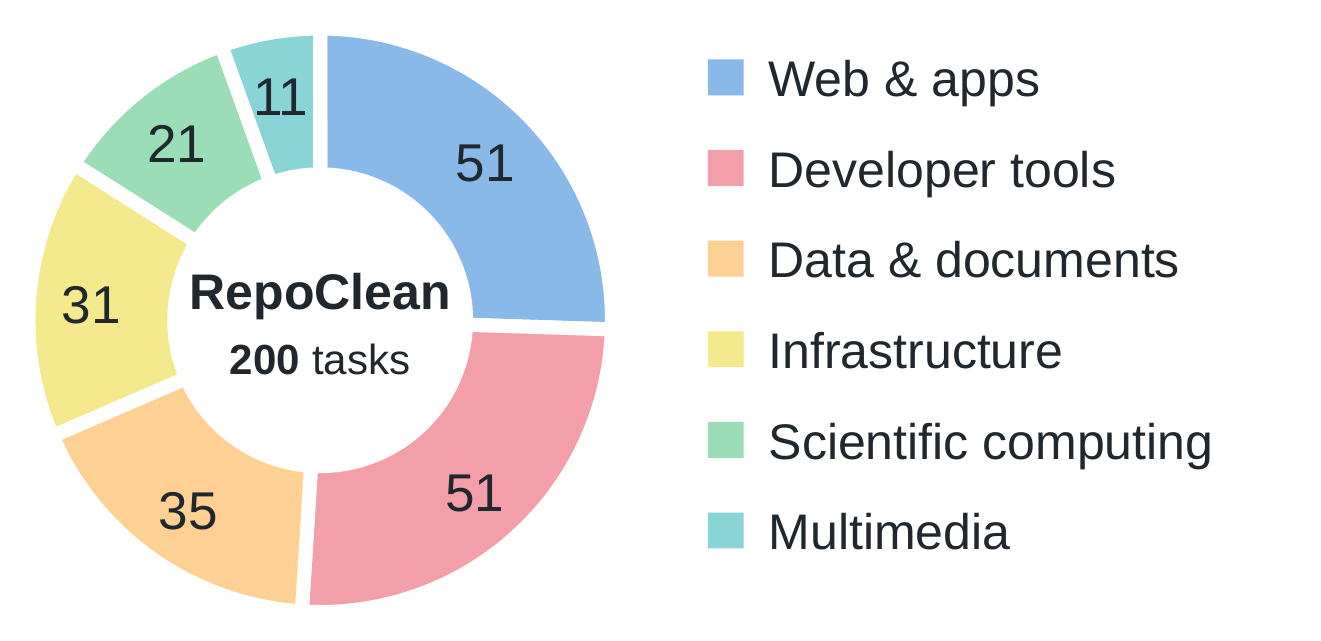}
    \caption{Distribution of tasks across six software domains in \name{}.}
    \label{fig:repository-distribution}
\end{figure}

\input{tables/dataset_statistics}

\subsection{Benchmark Composition}

\paragraph{Full Set.}
The pipeline retains candidates with 100--600 lines of historical code involving 10--200 functions or methods. Considering repository coverage, we curate 200 validated tasks across 34 Python repositories to form \name{}. Figure~\ref{fig:repository-distribution} shows their distribution across six domains, including web development, developer tools, and scientific computing. Appendix~\ref{app:benchmark-composition} provides a detailed breakdown of the benchmark. As summarized in Table~\ref{tab:dataset-statistics}, gold patches remove an average of 256.85 lines across 15.62 files. This scope reflects maintenance scenarios where obsolete code is spread across multiple files.

\paragraph{Lite Set.}
To reduce evaluation cost, we select 40 tasks from the full set to form \name{} Lite. Lite covers 27 repositories and all six domains. Performance trends across models are broadly consistent between Lite and Full.

\section{Experiments}
\label{sec:experiments}

\input{tables/main_results}

\subsection{Experimental Setup}
\label{sec:experimental-setup}

\paragraph{Models and frameworks.}
The evaluation includes seven models: GPT-6 Astra \citep{openai2026gpt6astra}, GPT-6.1 Sol \citep{openai2026gpt61sol}, DeepSeek-V4.1-Flash \citep{deepseekai2026deepseekv41flashpushinglimitskv}, GLM-5.3 \citep{glm5team2026glm5vibecodingagentic}, Kimi K3 \citep{kimiteam2026kimik3openfrontier}, Qwen3.8-27B \citep{qwen38}, and Qwen3.6-35B-A3B \citep{qwen36_35b_a3b}. All models use mini-swe-agent 2.4.6 \citep{NEURIPS2024_5a7c9475}, with Harbor \citep{Harbor_Framework} managing execution and final validation. Appendix~\ref{app:agent-evaluation-details} provides model configurations and runtime settings.

\paragraph{Evaluation.}
Evaluation comprises three stages: environment preparation, agent execution, and final validation. During environment preparation, an isolated Docker container with 2 CPUs and 8 GB of memory is initialized for each run within a 60-minute limit. During agent execution, each model runs under a 60-minute limit covering model calls and tool execution, without a step limit. Agents cannot access final validation tests or perform online searches or downloads. During final validation, the resulting patch is applied to the initial code state in a separate container and tested under a 30-minute limit. Reported metrics include Resolved, Recovery, and Coverage (Section~\ref{sec:evaluation-metrics}), along with task-averaged input and output token counts and interaction steps.

\subsection{Main Results}
\label{sec:main-results}

Table~\ref{tab:main-results} reports the results of seven models on \name{} Full and Lite, where the highest resolution rates reach only 23\% and 20\%, respectively. Even when the functionality required by the current version is already implemented, agents struggle to complete cleanup reliably. Among the 1,201 unresolved runs on \name{}, only 2.1\% have no deleted lines in their patches, and 12.4\% achieve deletion coverage of at least 90\%. These results show that making deletions, even with high deletion coverage, does not guarantee task resolution, highlighting repository-level cleanup as a remaining capability gap. Across Lite and Full, model rankings are broadly consistent, with Spearman correlations of 0.96 for Coverage, 0.82 for Recovery, and 0.79 for Resolved.

The results also reveal substantial differences in interaction steps across models. On \name{}, DeepSeek-V4.1-Flash averages approximately 8.6 times as many interaction steps as GPT-6 Astra, despite a slightly lower resolution rate. Together, these findings highlight the need to improve both cleanup reliability and execution efficiency.

\begin{figure}[!t]
    \centering
    \includegraphics[width=\columnwidth]{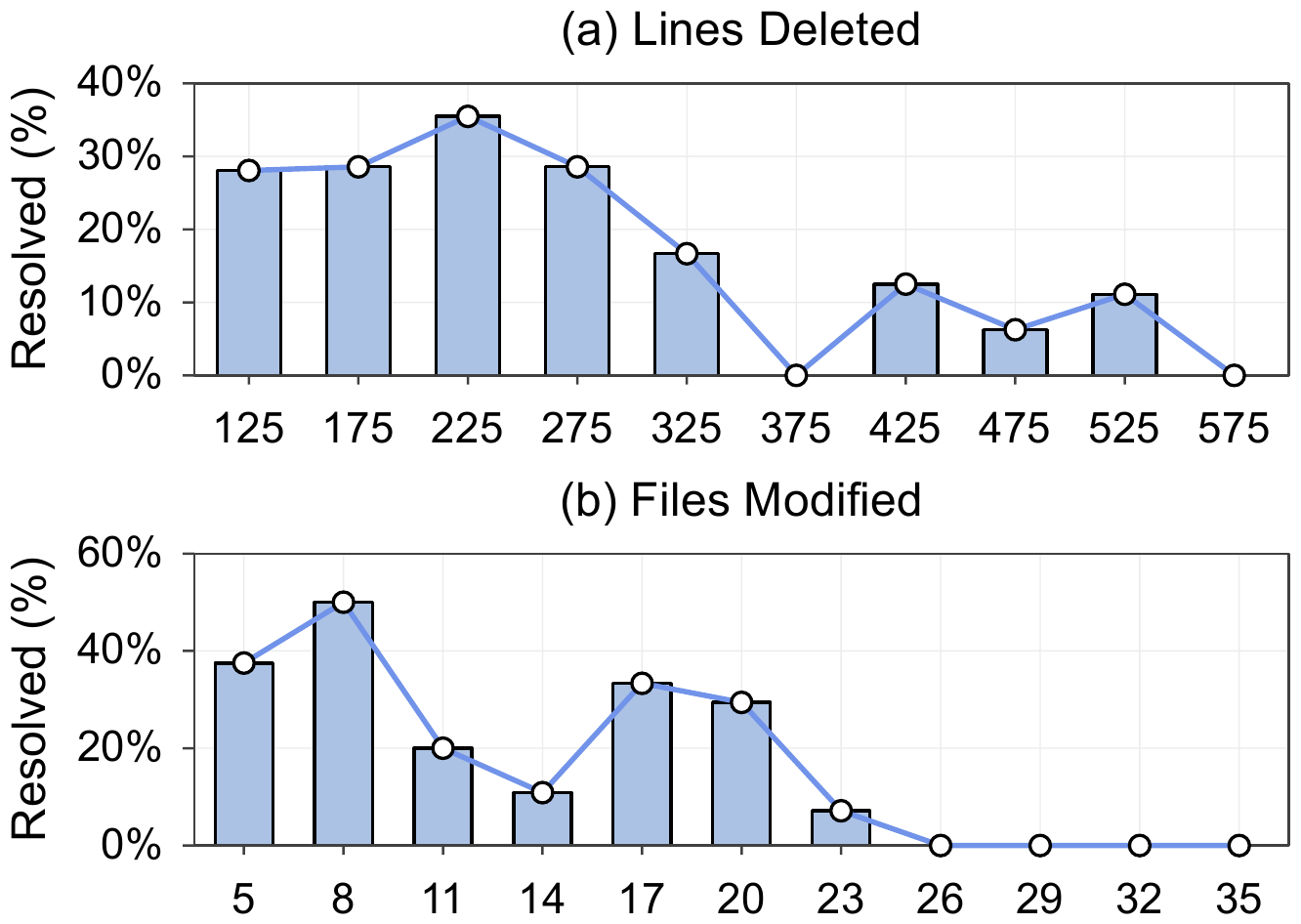}
    \caption{Cleanup scope and resolution rates for GPT-6 Astra on \name{}. Tasks are grouped by (a) lines deleted and (b) files modified by the gold patch, using bins of 50 lines and three files, respectively. Bars show Resolved (\%) per bin; ticks mark bin centers.}
    \label{fig:task-characteristics}
\end{figure}

\begin{figure*}[t]
    \centering
    \includegraphics[width=\textwidth]{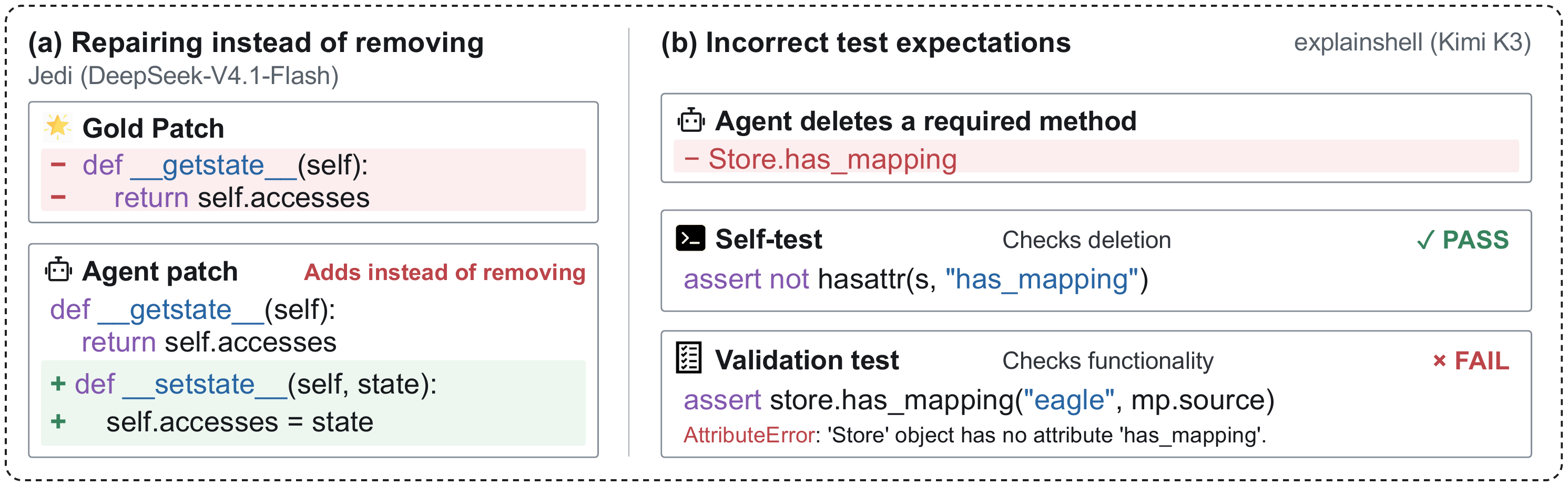}
    \caption{Examples of cleanup errors and misleading self-tests. (a) DeepSeek retains an old method that the gold patch removes and adds a supporting method instead. (b) Kimi deletes a method still required by the current version. Its self-test passes by checking only that the method is absent, while a previously passing validation test fails when calling it. Code and logs are shown as excerpts.}
    \label{fig:agent-patch-cases}
\end{figure*}

\subsection{Ablation and Analysis}
\label{sec:ablation-studies}

\subsubsection{Ablation Studies}
\label{sec:ablation-experiments}

The ablation studies evaluate GPT-6.1 Sol, DeepSeek-V4.1-Flash, and Qwen3.8-27B on \name{} Lite. Only the task instructions vary (Appendix~\ref{app:evaluation-prompts}), while the initial code, environments, and time budgets remain unchanged.

\paragraph{Effect of file-location hints.}
The cleanup instruction is supplemented with paths to files containing historical code, without identifying the obsolete code within each file. Table~\ref{tab:ablation-file-hints} shows that these hints yield uneven benefits across models. GPT-6.1 Sol improves on all three metrics, with Resolved rising from 15.0\% to 27.5\%. For DeepSeek, however, higher Recovery and Coverage do not translate into a higher resolution rate. These results suggest that reliable cleanup requires reasoning about code dependencies and functional behavior across the repository, beyond identifying the files to edit.

\input{tables/ablation_file_hints}

\input{tables/ablation_cleanup_instruction}

\paragraph{Effect of cleanup instructions.}
To assess autonomous cleanup, the explicit cleanup instruction is replaced with a general request to repair functionality. As shown in Table~\ref{tab:ablation-cleanup-instruction}, Coverage drops by approximately 28--37 percentage points across all three models, while Recovery changes comparatively little. These results reveal limitations in current models' ability to autonomously clean up obsolete code during functional repair.

\subsubsection{Task Characteristics}
\label{sec:task-characteristics}

\paragraph{Deleted lines.}
Figure~\ref{fig:task-characteristics}(a) shows that GPT-6 Astra's resolution rate on \name{} declines overall as gold patches delete more lines. This pattern suggests that reliable cleanup must scale from recognizing individual remnants to identifying the full extent of obsolete code.

\begin{figure}[!b]
    \centering
    \includegraphics[width=\columnwidth]{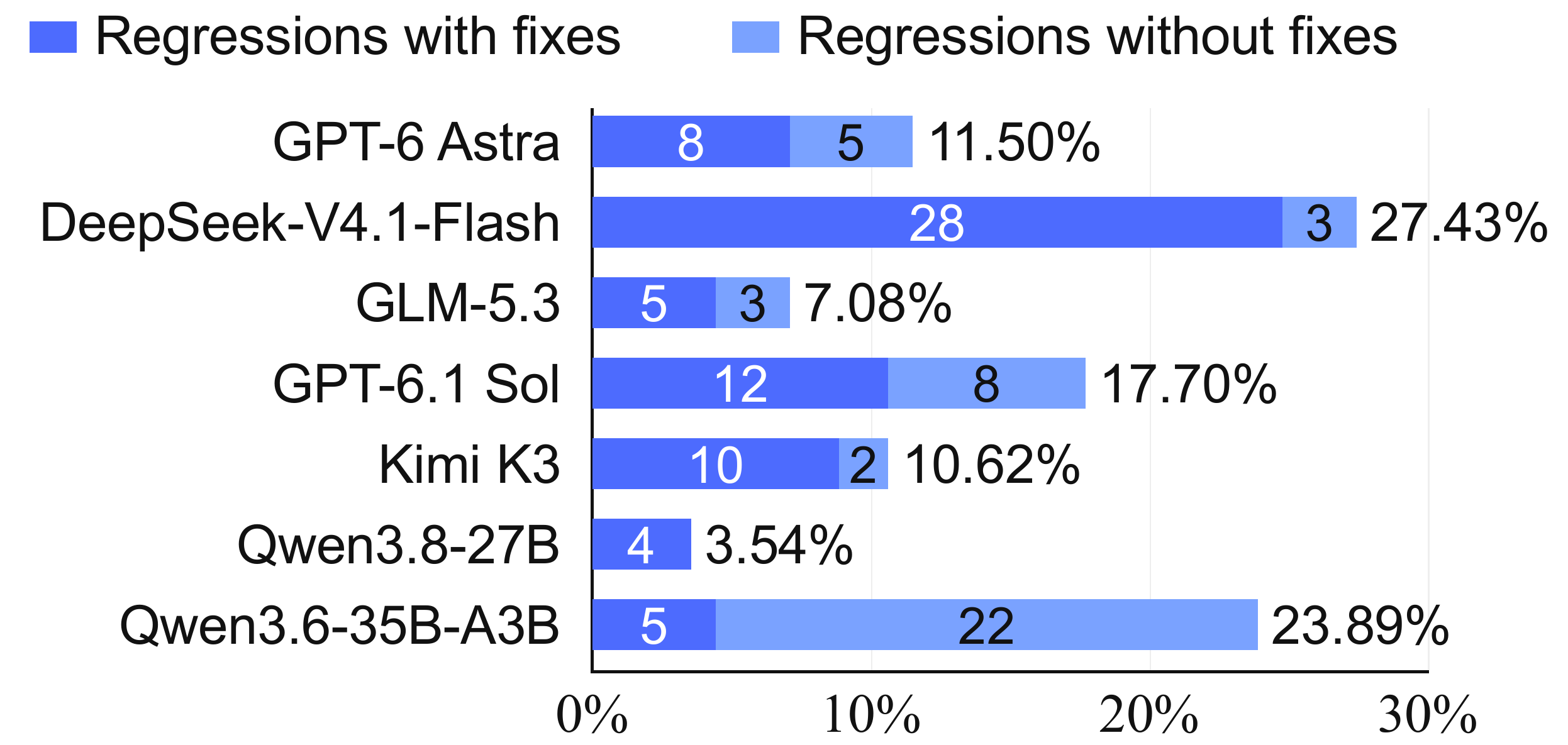}
    \caption{Functional regressions during code cleanup. Results cover tasks in \name{} with initially passing validation test files. A regression occurs when an initially passing test file fails after editing; a fix occurs when an initially failing file passes. Numbers within bars indicate task counts; percentages at bar ends indicate the overall regression rate.}
    \label{fig:functional-regressions}
\end{figure}

\begin{figure*}[t]
    \centering
    \includegraphics[width=\textwidth]{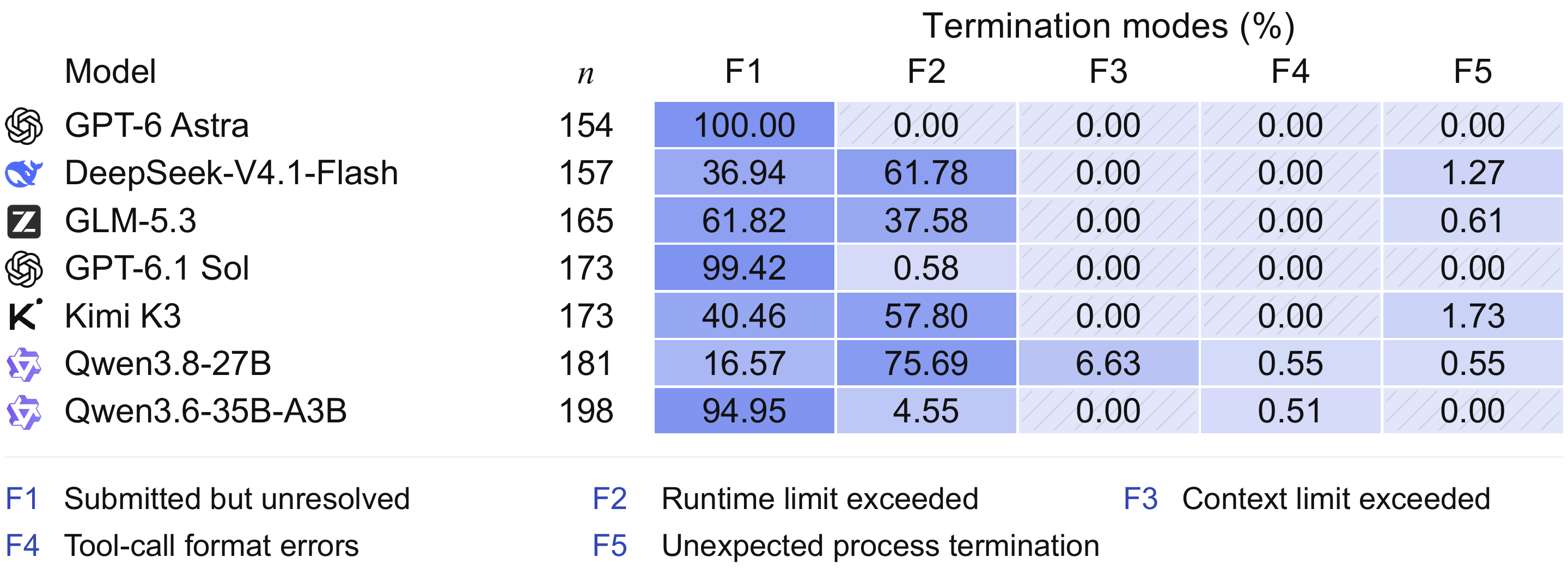}
    \caption{Termination modes of unresolved runs. Each row reports category percentages among a model's unresolved runs on \name{}; \(n\) denotes the number of such runs. Colors use a nonlinear scale; hatched cells indicate zero occurrences.}
    \label{fig:termination-outcomes}
\end{figure*}

\paragraph{Modified files.}
Figure~\ref{fig:task-characteristics}(b) shows a similar downward trend as gold patches modify more files. With cleanup spread across more files, a deletion must be assessed against callers and shared interfaces beyond its immediate location. The trend highlights the need to coordinate removals across file boundaries, using repository-wide dependencies to distinguish obsolete code from functionality still required by the current implementation.

\subsection{Agent Behavior and Failure Analysis}
\label{sec:agent-behavior-failure-analysis}

\paragraph{Agent behavior.}
Even with explicit cleanup instructions, agents may retain obsolete code or delete code still required by the current version. In the Jedi task in Figure~\ref{fig:agent-patch-cases}(a), DeepSeek retains an obsolete method for saving object state and adds a method for restoring it, while the gold patch removes the obsolete method. In the explainshell task in Figure~\ref{fig:agent-patch-cases}(b), Kimi deletes a required interface and writes a self-test that checks for its absence. This self-test passes, but a previously passing validation test file fails because the interface is missing. These cases show that agents can misjudge what to retain or remove, and their self-tests may reinforce rather than expose these errors.

\begingroup
\widowpenalty=10000
\paragraph{Functional regressions.}
We further examine whether agents restore affected functionality without introducing regressions. Figure~\ref{fig:functional-regressions} reports regressions on the 113 tasks in \name{} with at least one validation test file that passes before cleanup, covering 56.5\% of the benchmark. In the remaining 87 tasks, the reintroduced code causes all validation test files to fail, leaving no initially passing files against which to assess regressions. All seven models cause previously passing files to fail, with regressions occurring in 115 runs. Of these, 72 also make at least one initially failing file pass, showing that repair and regression can coexist within a single run. Local repair progress therefore does not guarantee that previously working functionality is preserved.
\par
\endgroup

\paragraph{Termination of unresolved runs.}
To extend the failure analysis to all unresolved runs, we summarize their termination modes in Figure~\ref{fig:termination-outcomes}. Submission without resolution and the 60-minute runtime limit account for approximately 98\% of the 1,201 unresolved runs on \name{}. Among unresolved runs, GPT-6 Astra and GPT-6.1 Sol almost always end with a submission, whereas DeepSeek-V4.1-Flash, Kimi K3, and Qwen3.8-27B more often reach the time limit. Trajectory inspection shows that some timed-out runs are still examining failure logs or attempting edits when execution stops. Context limits, repeated tool-call format errors, and unexpected process termination account for the remainder. Together, these findings highlight the need to identify necessary changes and verify their effects within the execution budget.

\begingroup
\brokenpenalty=10000
\section{Conclusion}

\name{} establishes autonomous repository-level code cleanup as a distinct evaluation target, with 200 tasks from 34 Python projects. Its automated pipeline turns real repository histories into verifiable cleanup tasks at scale, creating reusable environments for agent evaluation. Experiments reveal a gap between functional repair and autonomous maintenance: agents struggle to remove obsolete code safely and do not consistently integrate cleanup into repair. \name{} provides a foundation for advancing coding agents toward autonomous software maintenance.

\par
\endgroup

\section*{Limitations}

\name{} currently focuses on Python repositories because its AST-based extraction pipeline is implemented for Python code. Extending the benchmark to other languages requires adapting code extraction to their syntax and validating the resulting tasks within their software ecosystems. Functional validation relies on repository tests, which provide executable evidence of correctness for the behaviors they cover. However, even when all tests pass, correctness in untested execution paths and usage scenarios requires further validation.

\bibliography{custom}

\clearpage
\appendix
\section{Historical Code Extraction Details}
\label{app:historical-code-extraction}

\begin{figure*}[!b]
\centering
\begingroup
\definecolor{fig5frame}{RGB}{133,141,151}
\definecolor{fig5header}{RGB}{48,52,58}
\definecolor{fig5ink}{RGB}{23,25,29}
\definecolor{fig5slate}{RGB}{89,97,106}
\definecolor{fig5red}{RGB}{188,68,73}
\definecolor{fig5delbg}{RGB}{252,238,239}
\definecolor{fig5kw}{RGB}{131,88,175}
\definecolor{fig5blue}{RGB}{40,101,163}
\lstdefinestyle{fig5code}{
    language=Python,
    basicstyle=\ttfamily\footnotesize\color{fig5ink},
    keywordstyle=\color{fig5kw},
    stringstyle=\color{fig5blue},
    commentstyle=\color{fig5slate},
    morecomment=[s][\color{fig5slate}]{"""}{"""},
    emph={_python2_dct_keys_to_unicode,pickle_dump},
    emphstyle=\color{fig5blue},
    columns=fixed,
    basewidth=0.51em,
    keepspaces=true,
    breaklines=false,
    showstringspaces=false,
    upquote=true,
    numbers=none,
    frame=none,
    aboveskip=0pt,
    belowskip=0pt,
    escapeinside={(*@}{@*)}
}
\begin{tcolorbox}[
    colback=white,
    colframe=fig5frame,
    colbacktitle=fig5header,
    coltitle=white,
    fonttitle=\bfseries\small,
    title={Pure-Deletion Diff Excerpt \hfill \normalfont\footnotesize\texttt{davidhalter/jedi}\enspace\textbar\enspace\texttt{jedi/\_compatibility.py}},
    arc=3pt,
    boxrule=0.6pt,
    left=7pt,
    right=7pt,
    top=5pt,
    bottom=5pt,
    toptitle=3pt,
    bottomtitle=3pt
]
\noindent\hspace*{4pt}{\ttfamily\footnotesize\color{fig5slate}@@ -403,24 +31,8 @@}\\[-1pt]
\begin{tcolorbox}[
    colback=fig5delbg,
    colframe=fig5delbg,
    arc=2pt,
    boxrule=0pt,
    left=4pt,
    right=6pt,
    top=2.5pt,
    bottom=2.5pt,
    before skip=3pt,
    after skip=3pt
]
\begin{lstlisting}[style=fig5code]
(*@\textcolor{fig5red}{-}@*)def _python2_dct_keys_to_unicode(data):(*@\hfill\textsf{\bfseries\scriptsize\textcolor{fig5red}{Fully deleted function}}@*)
(*@\textcolor{fig5red}{-}@*)    """
(*@\textcolor{fig5red}{-}@*)    Python 2 stores object __dict__ entries as bytes, not unicode, correct it
(*@\textcolor{fig5red}{-}@*)    here. Python 2 can deal with both, Python 3 expects unicode.
(*@\textcolor{fig5red}{-}@*)    """
(*@\textcolor{fig5red}{-}@*)    if isinstance(data, tuple):
(*@\textcolor{fig5red}{-}@*)        return tuple(_python2_dct_keys_to_unicode(x) for x in data)
(*@\textcolor{fig5red}{-}@*)    elif isinstance(data, list):
(*@\textcolor{fig5red}{-}@*)        return list(_python2_dct_keys_to_unicode(x) for x in data)
(*@\textcolor{fig5red}{-}@*)    elif hasattr(data, '__dict__') and type(data.__dict__) == dict:
(*@\textcolor{fig5red}{-}@*)        data.__dict__ = {unicode(k): v for k, v in data.__dict__.items()}
(*@\textcolor{fig5red}{-}@*)    return data
\end{lstlisting}
\end{tcolorbox}
\begin{tcolorbox}[
    colback=white,
    colframe=white,
    arc=0pt,
    boxrule=0pt,
    left=4pt,
    right=6pt,
    top=0.5pt,
    bottom=0.5pt,
    before skip=1pt,
    after skip=1pt
]
\begin{lstlisting}[style=fig5code]
 def pickle_dump(data, file, protocol):
     try:
\end{lstlisting}
\end{tcolorbox}
\begin{tcolorbox}[
    colback=fig5delbg,
    colframe=fig5delbg,
    arc=2pt,
    boxrule=0pt,
    left=4pt,
    right=6pt,
    top=2.5pt,
    bottom=2.5pt,
    before skip=1pt,
    after skip=1pt
]
\begin{lstlisting}[style=fig5code]
(*@\textcolor{fig5red}{-}@*)        if not is_py3:(*@\hfill\textsf{\bfseries\scriptsize\textcolor{fig5red}{Internal fragment in \texttt{pickle\_dump}}}@*)
(*@\textcolor{fig5red}{-}@*)            data = _python2_dct_keys_to_unicode(data)
\end{lstlisting}
\end{tcolorbox}
\begin{tcolorbox}[
    colback=white,
    colframe=white,
    arc=0pt,
    boxrule=0pt,
    left=4pt,
    right=6pt,
    top=0.5pt,
    bottom=0pt,
    before skip=1pt,
    after skip=0pt
]
\begin{lstlisting}[style=fig5code]
         pickle.dump(data, file, protocol)
\end{lstlisting}
\end{tcolorbox}
\end{tcolorbox}
\endgroup
\caption{Excerpt of the pure-deletion diff in \texttt{davidhalter/jedi} (\texttt{jedi/\_compatibility.py}, \texttt{0c618a4456c4} $\to$ \texttt{ef5d36112106}). Lines prefixed with \texttt{-} denote historical code deleted from \(C_k\) and extracted by the pipeline, including the entire function \texttt{\_python2\_dct\_keys\_to\_unicode} and a two-line internal fragment inside the retained function \texttt{pickle\_dump}.}
\label{fig:extraction-example}
\end{figure*}

\paragraph{Historical version sampling.}
For each repository, we use the latest commit on the default branch at collection time as the target version \(C_0\). We traverse its history along the first-parent chain, following the first parent whenever a merge commit is encountered. We sample every commit at distances of 1--100 commits from the target version and every 25 commits at distances of 125--750 commits. Each sampled version \(C_k\) is paired with the same target version \(C_0\). For repositories with shorter histories, sampling is limited to the available commits.

\paragraph{Code extraction and matching.}
We retain only Python source files that exist at the same path in both \(C_k\) and \(C_0\), change between the two versions, and parse cleanly with Python's abstract syntax tree (AST) parser in both versions. We exclude tests, documentation, examples, build and cache directories, CI workflows, and environment or packaging configuration files.

For each retained file, we extract code exclusively from pure-deletion diff hunks between \(C_k\) and \(C_0\) that contain deleted lines and no added lines, discarding any hunk that includes additions. Restricting extraction to pure deletions ensures that reinserting the extracted code only appends historical lines without overwriting current implementations in \(C_0\).

Using ASTs for both versions, we determine the source line range of each class, function, and method definition and apply two extraction rules, skipping fragments consisting solely of blank lines or comments:
\begin{itemize}
    \item \textbf{Fully deleted objects.} If a class, function, or method lies entirely within a pure-deletion range in \(C_k\), we extract it as a complete unit. When nested definitions occur, we select the outermost enclosing object first so that methods inside a fully deleted class are not extracted twice.
    \item \textbf{Internal fragments.} Remaining deleted lines not covered by a fully deleted object are assigned to their innermost enclosing function or method in \(C_k\); module-level and class-body statements outside any function or method are ignored. For each resulting fragment, we locate the corresponding definition in \(C_0\) by matching its kind (function or method) and qualified name, and retain the fragment only if the diff-indicated insertion line falls within that target definition's line range in \(C_0\).
\end{itemize}

\paragraph{Size measurement.}
We retain candidate version pairs with 100--600 lines of extracted code across 10--200 functions or methods. The line count sums all extracted source lines across both fully deleted objects and internal fragments, and Table~\ref{tab:extraction-counting-rules} summarizes how functions and methods are counted.

\begin{table}[!t]
\centering
\begingroup
\small
\setlength{\tabcolsep}{5pt}
\renewcommand{\arraystretch}{1.18}
\renewcommand{\tabularxcolumn}[1]{m{#1}}
\begin{tabularx}{\columnwidth}{>{\raggedright\arraybackslash}m{0.36\columnwidth}>{\raggedright\arraybackslash}X}
\toprule
\textbf{Extracted content} & \textbf{Counting rule} \\
\midrule
\rowcolor[gray]{0.95}
Entire function or method & Counts as 1 each. \\
Internal fragments & Enclosing function or method counts once, regardless of fragment count. \\
\rowcolor[gray]{0.95}
Entire class & Each member function or method counts as 1; the class itself is not counted. \\
\bottomrule
\end{tabularx}
\endgroup
\caption{Function and method counting rules for extracted historical code.}
\label{tab:extraction-counting-rules}
\end{table}

\paragraph{Illustrative example.}
Figure~\ref{fig:extraction-example} illustrates both extraction rules on a real version pair from \texttt{davidhalter/\allowbreak jedi} (\texttt{jedi/\_\allowbreak compatibility.py}), where \(C_k\) (\texttt{0c618a4456c4}) precedes \(C_0\) (\texttt{ef5d36112106}) by 250 commits. When the repository dropped Python~2 compatibility, the helper function \texttt{\_python2\_\allowbreak dct\_keys\_\allowbreak to\_unicode} was deleted in full, and a two-line compatibility branch inside the retained function \texttt{pickle\_dump} was removed. The extractor retains the helper as a fully deleted function (one unit) and matches the two-line branch to \texttt{pickle\_dump} in \(C_0\) after verifying that its insertion position falls within the target function body (one unit). In contrast, deleted module-level statements in the same file (such as the definition of \texttt{is\_py3}) lie outside any function or method and are ignored.

\section{Environment Construction Details}
\label{app:environment-construction}

This appendix supplements Section~\ref{sec:task-construction} with details of the two-stage environment construction workflow, the unified test script interface, validation test set selection, runtime settings, and agent prompts. For each repository's target version \(C_0\), the pipeline prepares a reusable runtime environment, a test execution script, and a fixed validation test set \(T\). All cleanup tasks generated from that version share the same environment, script, and test set.

\subsection{Two-Stage Construction Workflow}
\label{app:env-workflow}

Environment construction consists of a planning stage and an iterative construction stage. Both stages operate with read-only access to the target repository checkout at \(C_0\) so that environment setup does not alter the source code.

\paragraph{Stage 1: Environment planning.}
A planning agent inspects the repository to assess whether an automated unit-test environment is feasible. It examines repository evidence in priority order, starting from manifests and lockfiles, followed by CI configurations, setup documentation, test runner configurations, helper scripts, and a small representative sample of tests, to infer the runtime family, dependency manager, build system, and test entrypoints. Using a curated catalog of prebuilt base images, the planning agent outputs a structured decision with three possible outcomes:
\begin{itemize}
    \item \texttt{ready}: The repository is feasible for automated environment construction and matches a base image in the catalog. The agent selects one base image and produces repository-specific guidance specifying the detected runtime and dependency manager, likely installation commands, test-file discovery rules, and single-file execution commands.
    \item \texttt{abandoned}: The repository lacks a meaningful unit-test corpus, relies primarily on integration or manual testing, or cannot be cleanly expressed under the unified test script contract. Processing terminates and the repository is excluded.
    \item \texttt{defect}: The repository appears feasible in principle, but the current base image catalog lacks the required runtime or base dependencies. The agent records the reason and proposes a base-image Dockerfile for catalog expansion.
\end{itemize}

\paragraph{Stage 2: Iterative construction and validation.}
When planning returns \texttt{ready}, a construction agent receives the repository metadata, selected base image references, and planning guidance to produce two artifacts. First, it writes a \texttt{Dockerfile} based on the selected image that builds a self-contained test environment bound to the live repository working tree rather than a detached packaged copy, ensuring that subsequent code edits take effect without rebuilding the image. Second, it writes \texttt{run\_script.sh} to implement the unified test script interface (Section~\ref{app:test-script-interface}).

After generating these artifacts, the construction agent calls the built-in \texttt{validate} tool on the host. The tool performs artifact schema checks, builds the Docker image, collects test files, and runs an initial sampled check (Section~\ref{app:val-test-selection}), returning the status, error information, and test statistics to the agent. Table~\ref{tab:validate-outcomes} summarizes the subsequent action for each return status. Whether a retry is allowed depends on the failure type and the number of remaining validation attempts.

\begin{table}[!t]
\centering
\begingroup
\small
\setlength{\tabcolsep}{5pt}
\renewcommand{\arraystretch}{1.18}
\renewcommand{\tabularxcolumn}[1]{m{#1}}
\begin{tabularx}{\columnwidth}{>{\raggedright\arraybackslash}m{0.31\columnwidth}>{\raggedright\arraybackslash}X}
\toprule
\textbf{Return status} & \textbf{Subsequent action} \\
\midrule
\rowcolor[gray]{0.95}
\texttt{retry} & The agent revises the Dockerfile or test script using the error feedback and calls \texttt{validate} again. \\
\texttt{smoke\_passed} & The agent stops editing, and the system proceeds to full validation (Section~\ref{app:val-test-selection}). \\
\rowcolor[gray]{0.95}
\texttt{failed} & Validation stops without further retries, and the repository is excluded. \\
\bottomrule
\end{tabularx}
\endgroup
\caption{Actions triggered by each return status of the \texttt{validate} tool.}
\label{tab:validate-outcomes}
\end{table}

\subsection{Unified Test Script Interface}
\label{app:test-script-interface}

Repositories vary in their test frameworks and directory layouts. To allow benchmark construction and subsequent evaluation to process results uniformly without repository-specific parsers, the generated \texttt{run\_script.sh} implements a two-action command-line interface that emits structured JSON:
\begin{itemize}
    \item \textbf{Test discovery (\texttt{collect}).} Discovers unit test files in the repository and outputs their paths relative to the repository root.
    \item \textbf{Single-file execution (\texttt{run}).} Executes one selected test file and reports its file-level status together with test-case counts. Even when test cases fail, the script still outputs valid JSON reporting the failure counts in \texttt{status} and \texttt{summary}, allowing the evaluator to distinguish test failures from script or environment errors.
\end{itemize}

Figure~\ref{fig:script-interface} shows both commands and sample JSON outputs from \texttt{aws/chalice}. When tests fail, \texttt{status} is reported as \texttt{"failed"} with nonzero \texttt{failed} or \texttt{errors} counts.

\begin{figure}[!t]
\centering
\begingroup
\definecolor{cardframe}{RGB}{133,141,151}
\definecolor{cardheader}{RGB}{48,52,58}
\definecolor{cardink}{RGB}{23,25,29}
\definecolor{cardslate}{RGB}{89,97,106}
\definecolor{cardblue}{RGB}{40,101,163}
\definecolor{cardkw}{RGB}{131,88,175}
\definecolor{cardgreen}{RGB}{54,131,93}
\lstdefinestyle{scriptcard}{
    basicstyle=\ttfamily\fontsize{6.7}{7.6}\selectfont\color{cardink},
    columns=fixed,
    basewidth=0.49em,
    keepspaces=true,
    breaklines=false,
    showstringspaces=false,
    upquote=true,
    numbers=none,
    frame=none,
    aboveskip=0pt,
    belowskip=0pt,
    escapeinside={(*@}{@*)}
}
\begin{tcolorbox}[
    colback=white,
    colframe=cardframe,
    colbacktitle=cardheader,
    coltitle=white,
    fonttitle=\bfseries\footnotesize,
    title={Unified Test Script Interface \hfill \normalfont\scriptsize\texttt{aws/chalice}},
    arc=3pt,
    boxrule=0.6pt,
    left=5pt,
    right=5pt,
    top=4pt,
    bottom=4pt,
    toptitle=2.5pt,
    bottomtitle=2.5pt
]
\begin{lstlisting}[style=scriptcard]
(*@\textcolor{cardslate}{\# Discover test files}@*)
(*@\textcolor{cardkw}{\$}@*) run_script.sh --action collect --out collect.json
{
  (*@\textcolor{cardblue}{"action"}@*): (*@\textcolor{cardgreen}{"collect"}@*),
  (*@\textcolor{cardblue}{"test\_files"}@*): [
    {(*@\textcolor{cardblue}{"path"}@*): "tests/unit/cli/test_cli.py"},
    {(*@\textcolor{cardblue}{"path"}@*): "tests/unit/deploy/test_deployer.py"},
    ...
  ]
}

(*@\textcolor{cardslate}{\# Run a single test file}@*)
(*@\textcolor{cardkw}{\$}@*) run_script.sh --action run \
    --target-selector tests/unit/cli/test_cli.py \
    --out run.json
{
  (*@\textcolor{cardblue}{"action"}@*): (*@\textcolor{cardgreen}{"run"}@*),
  (*@\textcolor{cardblue}{"status"}@*): (*@\textcolor{cardgreen}{"passed"}@*),
  (*@\textcolor{cardblue}{"summary"}@*): {
    (*@\textcolor{cardblue}{"collected"}@*): 2, (*@\textcolor{cardblue}{"passed"}@*): 2,
    (*@\textcolor{cardblue}{"failed"}@*): 0, (*@\textcolor{cardblue}{"errors"}@*): 0, (*@\textcolor{cardblue}{"skipped"}@*): 0
  }
}
\end{lstlisting}
\end{tcolorbox}
\endgroup
\caption{Example commands and JSON outputs of the unified \texttt{run\_script.sh} interface on \texttt{aws/chalice}.}
\label{fig:script-interface}
\end{figure}

\subsection{Validation Test Set Selection}
\label{app:val-test-selection}

\paragraph{Initial check.}
We randomly sample up to 10 collected test files for execution. If fewer than 10 files are available, we execute all of them. A fixed random seed is used within each environment construction run, so repeated checks use the same sample when the test file list remains unchanged. The initial check passes (returning \texttt{smoke\_passed}) if the sample is nonempty and at least half of the sampled files pass.

\paragraph{Full test execution.}
Once the initial check passes, the system runs all collected test files individually on the target version \(C_0\). Each file is executed in a newly started container to reduce interference between test executions.

A test file is considered to pass only if at least one test case passes and no test case fails or encounters an execution error. Files in which all test cases are skipped are excluded from the passing set. The files meeting these criteria form the fixed validation test set \(T\), which is saved together with the Dockerfile, test execution script, and execution results for subsequent task construction. Repositories are excluded without further retries if a usable runtime environment cannot be obtained, full validation fails, or no test files pass.

\lstdefinestyle{environmentprompt}{
    basicstyle=\ttfamily\footnotesize,
    columns=fullflexible,
    keepspaces=true,
    breaklines=true,
    breakatwhitespace=false,
    showstringspaces=false,
    upquote=true,
    numbers=none,
    frame=single,
    framerule=0.4pt,
    framesep=3pt,
    xleftmargin=4pt,
    xrightmargin=4pt
}

\lstdefinestyle{promptcardlisting}{
    basicstyle=\ttfamily\fontsize{7.0}{7.9}\selectfont,
    columns=fullflexible,
    keepspaces=true,
    breaklines=true,
    breakatwhitespace=true,
    breakindent=10pt,
    showstringspaces=false,
    upquote=true,
    numbers=none,
    frame=none,
    aboveskip=0pt,
    belowskip=0pt
}

\definecolor{promptcardheader}{RGB}{48,52,58}
\definecolor{promptcardbg}{RGB}{255,253,244}
\definecolor{promptcardframe}{RGB}{133,141,151}

\subsection{Runtime Settings}

Both environment planning and environment construction use GPT-5.6 Luna. Table~\ref{tab:environment-construction-settings} summarizes the runtime limits and timeout settings.

\newpage

\noindent\begin{minipage}[t]{\columnwidth}
\makeatletter\def\@captype{table}\makeatother
\vspace{0pt}
\centering
\begingroup
\small
\setlength{\tabcolsep}{5pt}
\renewcommand{\arraystretch}{1.08}
\begin{tabularx}{\columnwidth}{>{\raggedright\arraybackslash}Xc}
\toprule
\textbf{Configuration item} & \textbf{Setting} \\
\midrule
\rowcolor[gray]{0.95}
Planning and construction model & GPT-5.6 Luna \\
Iteration limit per agent & 150 \\
\rowcolor[gray]{0.95}
Time limit per agent & 120 minutes \\
Max validation attempts & 10 \\
\rowcolor[gray]{0.95}
Initial-check sample size & Up to 10 files \\
Docker image build timeout & 40 minutes \\
\rowcolor[gray]{0.95}
Test collection timeout & 10 minutes \\
Per-test-file timeout & 10 minutes \\
\rowcolor[gray]{0.95}
Full validation timeout & 120 minutes \\
\bottomrule
\end{tabularx}
\endgroup
\caption{Environment construction settings.}
\label{tab:environment-construction-settings}
\end{minipage}

\begin{figure*}[!b]
\centering
\begin{tcolorbox}[
    colback=promptcardbg,
    colframe=promptcardframe,
    colbacktitle=promptcardheader,
    coltitle=white,
    fonttitle=\bfseries\small,
    title=Environment Planning Prompt (Excerpt),
    arc=3pt,
    boxrule=0.6pt,
    left=6pt,
    right=6pt,
    top=4pt,
    bottom=4pt,
    toptitle=3pt,
    bottomtitle=3pt
]
\begin{lstlisting}[style=promptcardlisting]
Your job is to inspect the repository, decide whether this repository is suitable for automated unit-test environment construction, choose the best base image from the provided catalog when possible, and write precise instructions for the later worker agent when the repository is feasible.

Workspace layout:
- the target repository checkout is at `$HOME/repo`
- do not modify any file under `$HOME/repo`
- write your final JSON result to `{{RESULT_FILE}}`

Repository metadata:
```json
{{REPOSITORY_JSON}}
```

Curated base image catalog:
```json
{{BASE_IMAGE_CATALOG_JSON}}
```

Decision outcomes:
1. `status = "ready"`
   - Use this only when the repository appears feasible for this automated unit-test environment workflow and one provided base image is suitable.
2. `status = "abandoned"`
   - Use this when the repository does not have a meaningful unit-test corpus, only has a few scattered test files, mainly relies on integration/manual testing, or would be very hard to express as the required `run_script.sh` collect/run contract.
3. `status = "defect"`
   - Use this when the repository appears theoretically feasible, but the provided base image catalog is inadequate.
   - You must provide `upd_dockerfile`, which is a complete Dockerfile for the base image you believe should be added to the catalog. This is a base image proposal, not the final worker Dockerfile for this repository.

Task:
1. Inspect repository evidence in this order: manifests and lockfiles, CI configuration, setup documentation, test runner configuration, helper scripts, then a small representative sample of tests only if needed.
2. Infer the likely runtime family, dependency system, build system, and test entrypoints from repository evidence.
3. Decide whether the correct outcome is `ready`, `abandoned`, or `defect`.
4. If the outcome is `ready`, choose exactly one base image from the provided catalog and produce repo-specific guidance for the later worker agent that will generate `Dockerfile` and `run_script.sh`.

- The inspected test sample is evidence for inferring collection and execution rules, not the worker's final collected file set.

Output JSON schema:
{
  "status": "ready | abandoned | defect",
  "selected_base_image_id": "required when status is ready",
  "guidance": "required when status is ready",
  "reason": "required when status is abandoned or defect",
  "upd_dockerfile": "required when status is defect"
}

Requirements:
- When `status` is `ready`, `guidance` must state the detected runtime, dependency manager, build system, and test framework.
- When `status` is `ready`, `guidance` must specify likely dependency installation commands.
- When `status` is `ready`, `guidance` must specify how the worker should discover unit test files.
- When `status` is `ready`, `guidance` must specify how the worker should run one selected unit test file.
\end{lstlisting}
\end{tcolorbox}
\caption{Excerpt of the prompt used by the environment planning agent.}
\label{fig:env-planning-prompt}
\end{figure*}

\newpage

\subsection{Agent Prompts}

Figures~\ref{fig:env-planning-prompt} and~\ref{fig:env-construction-prompt} reproduce key excerpts from the environment planning and environment construction prompts, covering the inputs, main requirements, expected outputs, and handling of validation feedback. Placeholders of the form \texttt{\{\{\ldots\}\}} are filled at runtime with the corresponding repository information, base image information, file paths, or planning results.

\begin{figure*}[!p]
\centering
\begin{tcolorbox}[
    colback=promptcardbg,
    colframe=promptcardframe,
    colbacktitle=promptcardheader,
    coltitle=white,
    fonttitle=\bfseries\small,
    title=Environment Construction Prompt (Excerpt),
    arc=3pt,
    boxrule=0.6pt,
    left=6pt,
    right=6pt,
    top=4pt,
    bottom=4pt,
    toptitle=3pt,
    bottomtitle=3pt
]
\begin{lstlisting}[style=promptcardlisting]
Your job is to build a runnable unit-test environment for the repository mounted at `/workspace/repo`.

Workspace rules:
- The repository checkout is at `/workspace/repo`.
- Do not modify files under `/workspace/repo`.
- Write the final environment artifacts to:
  - `{{DOCKERFILE_PATH}}`
  - `{{RUN_SCRIPT_PATH}}`

Repository metadata:
```json
{{REPOSITORY_JSON}}
```

Selected base image summary:
{{SELECTED_BASE_IMAGE_SUMMARY}}

Reference files for the selected base image:
- Dockerfile reference: `{{SELECTED_BASE_IMAGE_DOCKERFILE_PATH}}`
- Metadata reference: `{{SELECTED_BASE_IMAGE_METADATA_PATH}}`

Top-level guidance from the planning pass:
{{PLANNER_GUIDANCE}}

Recommended working order:
1. Read the planning guidance and extract the intended install, collect, and run strategy.
2. Do a focused verification pass on the repository and selected base image references.
   - Confirm the runtime, dependency manager, test framework, and repo-native commands that matter for implementation.
   - Reuse preinstalled tooling, mirrors, and package-manager configuration when possible.
3. Narrow or correct the plan only where repository evidence disagrees.
   - Prefer deterministic, lockfile-driven installs.
   - Add only the OS packages and runtime steps that are still missing.
4. Design `run_script.sh` before finalizing the environment.
   - Decide how `collect` will discover unit-test files.
   - Decide how `run --target-selector <file>` will execute exactly one collected file.
   - Ensure ordinary failing tests still produce valid JSON.
5. Write `Dockerfile` and `run_script.sh`.
6. Self-check the artifact contract.

Your validate workflow:
- A custom tool named `{{VALIDATE_TOOL_NAME}}` is available.
- When you believe the current `Dockerfile` and `run_script.sh` are ready, call `{{VALIDATE_TOOL_NAME}}`.
- Pass the container paths of the files you want validated.
- The tool will run schema checks, build, collect, and smoke validation.
- If the tool returns `result = "retry"`, inspect the returned feedback, revise the files in place, and call `{{VALIDATE_TOOL_NAME}}` again.
- If the tool returns `result = "smoke_passed"`, stop editing the artifacts. The host will checkpoint the worker, end the agent run, and continue full validation outside the agent time budget.
- If the tool returns `result = "failed"`, no more useful retry path remains. Finish with a short failure summary.

Artifact contract:
1. `Dockerfile` must start with `FROM {{SELECTED_BASE_IMAGE_REF}}`.
2. `Dockerfile` must build a self-contained runnable test environment for `/workspace/repo`.
3. `run_script.sh` must support:
   - `--action collect --out <path>`
   - `--action run --target-selector <test-file> --out <path>`
4. `collect` must write JSON with:
   - `action: "collect"`
   - `test_files: [{"path": "tests/..."}, {"path": "pkg/tests/unit/test_a.py", "target_selector": "unit/test_a.py"}]`
5. `run` must write JSON with:
   - `action: "run"`
   - `status`
   - `summary.collected`
   - `summary.passed`
   - `summary.failed`
   - `summary.errors`
   - `summary.skipped`

Validator execution details you must satisfy:
- For `--action run`, ordinary test failures must still produce valid JSON. Encode test failures in the JSON `status` and `summary` fields.

Requirements:
- Build the environment so a future coding agent can modify `/workspace/repo` and re-run tests against the live working tree without rebuilding the image.
- Prefer install modes that keep the runtime bound to the current repository checkout instead of a detached packaged copy.
\end{lstlisting}
\end{tcolorbox}
\caption{Excerpt of the prompt used by the environment construction agent.}
\label{fig:env-construction-prompt}
\end{figure*}

\clearpage

\section{Task Assembly and Validation Details}
\label{app:task-assembly-validation}

This appendix supplements Section~\ref{sec:task-construction} with details on historical code reinsertion and deduplication, workspace preparation, and task validation.

\subsection{Reinsertion, Gold Patch, and Deduplication}

For each candidate version pair that passes size filtering, we insert the extracted objects and internal fragments into the target version \(C_0\) in reverse line order within each file so that earlier insertions do not shift subsequent insertion positions, producing a reinsertion patch relative to \(C_0\).

When multiple historical commits from the same repository produce byte-identical reinsertion patches on \(C_0\), the pipeline deduplicates them by patch content and retains only the candidate closest to \(C_0\) (smallest commit offset). From the resulting initial task state, the pipeline constructs the reverse deletion-only diff back to \(C_0\) as the task's gold patch.

\subsection{Workspace Preparation and Leakage Prevention}

To prevent agents from exploiting original Git history, file timestamps, or held-out validation tests, the workspace provided to the agent undergoes three leakage-prevention steps during task preparation:
\begin{itemize}
    \item \textbf{Hiding validation tests.} All files belonging to the fixed validation test set \(T\) are removed from the agent-visible working tree. Snapshots of these files are stored separately and restored only inside an isolated container during final evaluation.
    \item \textbf{Single-commit repository reconstruction.} The original \texttt{.git} directory is removed, and a new single-commit repository on \texttt{main} is initialized from a fresh copy of the working tree containing only the task's initial code state, with no Git remotes or historical objects.
    \item \textbf{Timestamp normalization.} The access and modification times of all visible files and directories in the working tree are set to a common fixed timestamp (\texttt{2000-01-01 00:00:00 UTC}) to remove timing clues about recently modified files.
\end{itemize}
After preparation, automated checks verify that the working tree is clean, contains exactly one commit, has no Git remotes or unreachable objects, retains no hidden validation test files, and has normalized timestamps across all visible paths.

\subsection{Functional Validation and Stability Verification}

\paragraph{Functional validation.}
Each deduplicated candidate undergoes two validation steps on the repository's fixed validation test set \(T\), with a 30-minute timeout per full-suite run. First, we rerun \(T\) on the unmodified target version \(C_0\) to confirm that all validation test files pass. Second, we run \(T\) on the initial task state after reinsertion, retaining the candidate only if test execution completes and at least one previously passing test file fails. Because the gold patch is the exact inverse of the reinsertion patch, applying it restores \(C_0\).

\paragraph{Stability verification.}
To filter out nondeterministic or order-sensitive tests, we perform three repeated validation passes covering the target version, the initial task state, and end-to-end recovery under the gold patch in the evaluation container, excluding candidates whose file-level pass or fail status fluctuates across runs. Across all three passes, all 200 tasks in the final benchmark consistently exhibit failing validation tests in the initial state, pass all validation test files after applying the gold patch, and maintain consistent file-level outcomes.

\section{Benchmark Composition}
\label{app:benchmark-composition}

Table~\ref{tab:repository-composition} groups the 34 Python repositories in \name{} across six software domains and reports their task counts in Full (200 tasks) and Lite (40 tasks). In Full, individual repositories contribute 1--17 tasks, with the largest share at 8.5\% (\texttt{Kozea/WeasyPrint}), preventing any single repository from dominating evaluation. Lite is a fixed subset of Full spanning 27 of the 34 repositories across all six domains and reuses the exact same initial code states, runtime environments, task instructions, and validation test sets as Full.

\begin{table*}[!t]
\centering
\small
\renewcommand{\arraystretch}{1.04}
\begin{tabularx}{\textwidth}{>{\raggedright\arraybackslash}p{0.28\textwidth}>{\raggedright\arraybackslash}Xrr}
\toprule
\textbf{Software domain} & \textbf{Repository} & \textbf{Full} & \textbf{Lite} \\
\midrule
\multirow[t]{9}{=}{Web \& apps} & aws/chalice & 5 & 1 \\
 & beeware/toga & 2 & 1 \\
 & django-extensions/django-extensions & 8 & 2 \\
 & encode/django-rest-framework & 7 & 1 \\
 & falconry/falcon & 5 & 1 \\
 & pallets/flask & 1 & 0 \\
 & pallets/werkzeug & 1 & 0 \\
 & pennersr/django-allauth & 13 & 3 \\
 & tornadoweb/tornado & 9 & 2 \\
\midrule
\multirow[t]{7}{=}{Developer tools} & PyGithub/PyGithub & 10 & 2 \\
 & davidhalter/jedi & 6 & 2 \\
 & google/yapf & 2 & 0 \\
 & idank/explainshell & 12 & 3 \\
 & pre-commit/pre-commit & 14 & 2 \\
 & pytest-dev/pytest & 4 & 0 \\
 & robotframework/robotframework & 3 & 1 \\
\midrule
\multirow[t]{6}{=}{Data \& documents} & Kozea/WeasyPrint & 17 & 2 \\
 & fonttools/fonttools & 5 & 1 \\
 & marshmallow-code/marshmallow & 2 & 1 \\
 & pdfminer/pdfminer.six & 6 & 1 \\
 & python-openxml/python-docx & 1 & 0 \\
 & wireservice/csvkit & 4 & 1 \\
\midrule
\multirow[t]{7}{=}{Infrastructure} & docker/docker-py & 3 & 1 \\
 & dpkp/kafka-python & 6 & 1 \\
 & fail2ban/fail2ban & 8 & 2 \\
 & kubernetes-client/python & 1 & 0 \\
 & magic-wormhole/magic-wormhole & 4 & 1 \\
 & patroni/patroni & 8 & 1 \\
 & redis/redis-py & 1 & 0 \\
\midrule
\multirow[t]{3}{=}{Scientific computing} & biopython/biopython & 4 & 1 \\
 & librosa/librosa & 13 & 3 \\
 & scikit-image/scikit-image & 4 & 1 \\
\midrule
\multirow[t]{2}{=}{Multimedia} & Zulko/moviepy & 9 & 1 \\
 & streamlink/streamlink & 2 & 1 \\
\midrule
\rowcolor[gray]{0.95}
\textbf{Total} & \textbf{34 repositories} & \textbf{200} & \textbf{40} \\
\bottomrule
\end{tabularx}
\caption{Task counts per repository in \name{} Full and Lite.}
\label{tab:repository-composition}
\end{table*}

\section{Agent Evaluation Details}
\label{app:agent-evaluation-details}

\subsection{Model Configurations}

Table~\ref{tab:evaluation-model-settings} lists the reasoning effort settings for the seven models. \texttt{default} denotes the default setting.

\begin{table}[!t]
\centering
\begingroup
\small
\setlength{\tabcolsep}{5pt}
\renewcommand{\arraystretch}{1.12}
\begin{tabularx}{\columnwidth}{>{\raggedright\arraybackslash}Xc}
\toprule
\textbf{Model} & \textbf{Reasoning effort} \\
\midrule
\rowcolor[gray]{0.95}
GPT-6 Astra & \texttt{xhigh} \\
GPT-6.1 Sol & \texttt{high} \\
\rowcolor[gray]{0.95}
DeepSeek-V4.1-Flash & \texttt{max} \\
GLM-5.3 & \texttt{max} \\
\rowcolor[gray]{0.95}
Kimi K3 & \texttt{max} \\
Qwen3.8-27B & \texttt{default} \\
\rowcolor[gray]{0.95}
Qwen3.6-35B-A3B & \texttt{default} \\
\bottomrule
\end{tabularx}
\endgroup
\caption{Reasoning effort settings for the evaluated models.}
\label{tab:evaluation-model-settings}
\end{table}

\subsection{Runtime Settings}

Table~\ref{tab:evaluation-runtime-settings} summarizes the settings for environment preparation, agent execution, container resources, and final validation.

Environment preparation and agent execution are timed separately, each with a 60-minute limit. Final validation takes place in a separate container under a 30-minute limit: the agent's patch is applied to the task's initial code state and evaluated against the validation tests prepared during benchmark construction.

Agents can run existing repository tests or write their own tests to check their changes. They cannot access the final validation test set or perform online searches or downloads.

\begin{table}[!t]
\centering
\begingroup
\small
\setlength{\tabcolsep}{5pt}
\renewcommand{\arraystretch}{1.12}
\begin{tabularx}{\columnwidth}{>{\raggedright\arraybackslash}Xc}
\toprule
\textbf{Setting} & \textbf{Value} \\
\midrule
\rowcolor[gray]{0.95}
Agent framework & mini-swe-agent 2.4.6 \\
Environment preparation limit & 60 minutes \\
\rowcolor[gray]{0.95}
Agent execution limit & 60 minutes \\
Agent interaction step limit & Unlimited \\
\rowcolor[gray]{0.95}
Terminal command timeout & 30 seconds \\
Final validation limit & 30 minutes \\
\rowcolor[gray]{0.95}
CPUs per container & 2 \\
Memory per container & 8,192 MB \\
\bottomrule
\end{tabularx}
\endgroup
\caption{Runtime settings for agents and evaluation containers.}
\label{tab:evaluation-runtime-settings}
\end{table}

\subsection{Scoring and Metric Computation}

When an agent run ends, whether through submission, reaching the 60-minute runtime limit, or terminating early due to context limits, tool-call format errors, or process interruption, the evaluation harness captures the working-tree changes against the initial task state as a patch for final validation. Deletion Coverage matches deleted non-blank source lines against the gold patch by file path, original line position in the initial task state, and line content, ignoring blank lines. If final validation exceeds the 30-minute limit, all validation test files are treated as failed (scoring Resolved and Recovery as 0), while Deletion Coverage is still computed from the captured patch.

\lstdefinestyle{evalpromptcard}{
    basicstyle=\ttfamily\fontsize{7.0}{7.9}\selectfont,
    columns=fullflexible,
    keepspaces=true,
    breaklines=true,
    breakatwhitespace=true,
    breakindent=10pt,
    showstringspaces=false,
    upquote=true,
    numbers=none,
    frame=none,
    aboveskip=0pt,
    belowskip=0pt,
    literate={’}{{\textquoteright}}1
}

\begin{figure*}[!b]
\centering
\begin{tcolorbox}[
    colback=promptcardbg,
    colframe=promptcardframe,
    colbacktitle=promptcardheader,
    coltitle=white,
    fonttitle=\bfseries\small,
    title=Cleanup Task Instruction,
    arc=3pt,
    boxrule=0.6pt,
    left=7pt,
    right=7pt,
    top=4pt,
    bottom=4pt,
    toptitle=3pt,
    bottomtitle=3pt
]
\begin{lstlisting}[style=evalpromptcard]
This repository is based on the target version, but remnants of obsolete implementations from an older version have been reintroduced. These remnants may be spread across multiple files, functions, and classes, and may involve independent behaviors.

These obsolete remnants may include, but are not limited to:

- branches, functions, classes, variables, or helper logic that are no longer used;
- old logic that duplicates or conflicts with the target-version implementation;
- old code that can still execute even though the target version no longer needs or expects it;
- historical logic that interferes with the target version’s control flow, data flow, or state management;
- obsolete imports, references, variables, or helper code left behind by the old implementation.

Please use the existing source code, call relationships, and available tests to identify and remove these obsolete implementations, bringing the affected behavior into alignment with the target version while preserving unaffected functionality.

Please follow these principles:

1. Use the target-version implementation already present in the repository as the basis for your work. Avoid redesigning or replacing existing implementations.
2. Remove the identified obsolete implementations and clean up related imports, references, variables, and helper code, taking care to preserve any parts still required by the target version.
3. Preserve code and behavior unrelated to these obsolete implementations, and avoid introducing unrelated changes.
4. Keep the final source changes clear and consistent with the surrounding code style and structure.

Repository and environment notes:

1. The project directory is `/workspace/repo`. Do not search other directories for another copy of the project’s source code.
2. External network access is unavailable in the current environment. Use the tools and dependencies already available, and do not attempt to access the network or install or update dependencies.
3. You may inspect and run the tests already present in the repository, and write additional tests as needed to help validate the changes.
4. Temporary diagnostic files may be placed under `/tmp`. When finished, remove temporary files and diagnostic artifacts, leaving only changes relevant to the task.
\end{lstlisting}
\end{tcolorbox}
\caption{Cleanup task instruction used in the main experiment.}
\label{fig:cleanup-task-prompt}
\end{figure*}

Token I/O is based on the cumulative input and output token counts saved at the end of each run. Steps is the number of agent response turns, including the final submission turn. For each model and experimental condition, we report arithmetic means over the corresponding task set.

\subsection{Evaluation Prompts}
\label{app:evaluation-prompts}

This section provides the task instructions used in the main experiment and the two ablation studies. Both ablations evaluate GPT-6.1 Sol, DeepSeek-V4.1-Flash, and Qwen3.8-27B on \name{} Lite. For each task, only the instructions vary across conditions, while the initial code, runtime environment, and time budget remain unchanged.

\paragraph{Cleanup task instruction.}
Figure~\ref{fig:cleanup-task-prompt} shows the cleanup instruction used in the main experiment, asking agents to autonomously identify and remove obsolete historical code, restore affected functionality, and preserve other functionality.

\paragraph{File-location hints.}
In the file-location hint ablation, we append paths to files containing historical code to the cleanup instruction without specifying which code within each file should be removed. Figure~\ref{fig:file-location-hints-prompt} shows an example from an actual task in the Chalice repository. The file list varies by task.

\paragraph{General repair instruction.}
To assess whether models can autonomously clean up historical code while repairing functionality, we replace the main experiment's explicit cleanup instruction with the general repair instruction shown in Figure~\ref{fig:general-repair-prompt}.

\makeatletter
\setlength{\@dblfptop}{0pt}
\setlength{\@dblfpsep}{18pt}
\setlength{\@dblfpbot}{0pt plus 1fil}
\makeatother

\begin{figure*}[!p]
\centering
\begin{tcolorbox}[
    colback=promptcardbg,
    colframe=promptcardframe,
    colbacktitle=promptcardheader,
    coltitle=white,
    fonttitle=\bfseries\small,
    title=File-Location Hints (Example: \texttt{aws/chalice}),
    arc=3pt,
    boxrule=0.6pt,
    left=7pt,
    right=7pt,
    top=4pt,
    bottom=4pt,
    toptitle=3pt,
    bottomtitle=3pt
]
\begin{lstlisting}[style=evalpromptcard]
File-location hint:

The reintroduced obsolete code is located in the following
repository-relative files:

- chalice/analyzer.py
- chalice/cli/factory.py
- chalice/deploy/deployer.py
- chalice/deploy/swagger.py
\end{lstlisting}
\end{tcolorbox}
\caption{Example file-location hints appended to the cleanup instruction in the file-location hint ablation.}
\label{fig:file-location-hints-prompt}
\end{figure*}

\begin{figure*}[!p]
\centering
\begin{tcolorbox}[
    colback=promptcardbg,
    colframe=promptcardframe,
    colbacktitle=promptcardheader,
    coltitle=white,
    fonttitle=\bfseries\small,
    title=General Repair Instruction,
    arc=3pt,
    boxrule=0.6pt,
    left=7pt,
    right=7pt,
    top=4pt,
    bottom=4pt,
    toptitle=3pt,
    bottomtitle=3pt
]
\begin{lstlisting}[style=evalpromptcard]
This repository is not working correctly. Please investigate what is going wrong and fix the problems so that it works correctly again.

Verify that your changes resolve the problems without breaking functionality that already works.

Please follow these principles:

1. Make the source-code changes needed to fix the problems.
2. Focus on the issues you identify and avoid unrelated edits.
3. Follow the surrounding code style and structure, keeping your changes clear and readable.

Repository and environment notes:

1. The project directory is `/workspace/repo`. Do not search other directories for another copy of the project’s source code.
2. External network access is unavailable in the current environment. Use the tools and dependencies already available, and do not attempt to access the network or install or update dependencies.
3. You may inspect and run the tests already present in the repository, and write additional tests as needed to help validate the changes.
4. Temporary diagnostic files may be placed under `/tmp`. When finished, remove temporary files and diagnostic artifacts, leaving only changes relevant to the task.
\end{lstlisting}
\end{tcolorbox}
\caption{General repair instruction used in the cleanup instruction ablation.}
\label{fig:general-repair-prompt}
\end{figure*}

\end{document}

%% file: tables/dataset_statistics.tex
\begin{table}[t]
\centering
\small
\renewcommand{\arraystretch}{1.12}
\begin{tabularx}{\columnwidth}{lX>{\columncolor{black!6}}rr}
\toprule
\textbf{Category} & \textbf{Statistic} & \textbf{Mean} & \textbf{Max} \\
\midrule
\multirow[t]{2}{*}{Gold patch} & \# Deleted lines & 256.85 & 573 \\
 & \# Modified files & 15.62 & 37 \\
\midrule
\multirow[t]{3}{*}{Test files} & \# Failed tests & 39.15 & 219 \\
 & \# Passed tests & 30.70 & 250 \\
 & \# Total & 69.85 & 267 \\
\bottomrule
\end{tabularx}
\caption{Per-task gold patch sizes and pre-cleanup test file counts in \name{}.}
\label{tab:dataset-statistics}
\end{table}

%% file: tables/main_results.tex
\begin{table*}[t]
\centering
\begingroup
\fontsize{9.5}{11.5}\selectfont
\setlength{\tabcolsep}{0.5pt}
\renewcommand{\arraystretch}{1.18}
\newcommand{\resultsep}{{\color[gray]{0.45}\vrule width 0.3pt}}
\begin{tabular*}{\textwidth}{@{\extracolsep{\fill}}c@{\hspace{2pt}\resultsep\hspace{2pt}}
>{\columncolor[gray]{0.95}}c c >{\columncolor[gray]{0.95}}c@{\hspace{5pt}}c c
@{\hspace{3.5pt}\resultsep\hspace{3.5pt}}
>{\columncolor[gray]{0.95}}c c >{\columncolor[gray]{0.95}}c@{\hspace{5pt}}c c@{}}
\toprule[0.8pt]
& \multicolumn{5}{c@{\hspace{3.5pt}\resultsep\hspace{3.5pt}}}{Full} & \multicolumn{5}{c}{Lite} \\
\cmidrule(lr){2-6}\cmidrule(l){7-11}
Model & \% Res. & \% Rec. & \% Cov. & Token I/O & \# Steps & \% Res. & \% Rec. & \% Cov. & Token I/O & \# Steps \\
\midrule[0.4pt]
GPT-6 Astra & 23.0 & 73.2 & 76.2 & 2.3M\,/\,16.9K & 33.9 & 15.0 & 72.2 & 75.5 & 1.8M\,/\,16.2K & 31.0 \\
DeepSeek-V4.1-Flash & 21.5 & 67.6 & 71.0 & 77.9M\,/\,168.9K & 293.0 & 20.0 & 64.0 & 69.7 & 74.5M\,/\,163.4K & 289.4 \\
GLM-5.3 & 17.5 & 72.5 & 70.7 & 28.6M\,/\,82.6K & 201.4 & 17.5 & 75.1 & 67.8 & 28.3M\,/\,83.6K & 202.2 \\
GPT-6.1 Sol & 13.5 & 67.4 & 70.2 & 1.9M\,/\,13.9K & 33.7 & 15.0 & 66.5 & 69.5 & 1.8M\,/\,14.1K & 32.8 \\
Kimi K3 & 13.5 & 68.7 & 65.5 & 19.2M\,/\,83.3K & 157.9 & 15.0 & 63.1 & 62.7 & 19.3M\,/\,83.8K & 160.0 \\
Qwen3.8-27B & 9.5 & 58.4 & 52.2 & 13.1M\,/\,75.0K & 120.1 & 10.0 & 58.4 & 50.0 & 12.6M\,/\,77.1K & 115.6 \\
Qwen3.6-35B-A3B & 1.0 & 23.1 & 14.3 & 7.6M\,/\,46.7K & 147.8 & 0.0 & 19.4 & 16.9 & 9.2M\,/\,50.7K & 166.2 \\

\bottomrule[0.8pt]
\end{tabular*}
\endgroup
\caption{Main results on \name{} Full and Lite, sorted by Full Resolved in descending order. Res., Rec., and Cov. denote Resolved, Recovery, and Coverage (\%). Token I/O and Steps are averaged over tasks.}
\label{tab:main-results}
\end{table*}

%% file: tables/ablation_file_hints.tex
\begin{table}[!ht]
\centering
\begingroup
\fontsize{10}{13}\selectfont
\setlength{\tabcolsep}{2pt}
\renewcommand{\arraystretch}{1.25}
\newcommand{\ablationsep}{{\color[gray]{0.45}\vrule width 0.3pt}}
\begin{tabular*}{\linewidth}{@{\extracolsep{\fill}}c@{\hspace{2pt}\ablationsep\hspace{2pt}}
c c c@{}}
\toprule[0.8pt]
& \multicolumn{3}{c}{With file paths} \\
\cmidrule(l){2-4}
Model & \% Res. & \% Rec. & \% Cov. \\
\midrule[0.4pt]
GPT-6.1 Sol & 27.5\,\textbf{(+12.5)} & 70.2\,(+3.7) & 84.6\,(+15.2) \\[2pt]
DS-V4.1-Flash & 17.5\,\textbf{(-2.5)} & 67.7\,\textbf{(+3.8)} & 77.1\,\textbf{(+7.3)} \\[2pt]
Qwen3.8-27B & 12.5\,(+2.5) & 44.9\,(-13.5) & 47.1\,(-2.9) \\[2pt]
\bottomrule[0.8pt]
\end{tabular*}
\endgroup
\caption{Effect of file-location hints on \name{} Lite, with changes relative to the no-hint baseline in Table~\ref{tab:main-results}.}
\label{tab:ablation-file-hints}
\end{table}

%% file: tables/ablation_cleanup_instruction.tex
\begin{table}[!b]
\centering
\begingroup
\fontsize{10}{13}\selectfont
\setlength{\tabcolsep}{2pt}
\renewcommand{\arraystretch}{1.25}
\newcommand{\ablationsep}{{\color[gray]{0.45}\vrule width 0.3pt}}
\begin{tabular*}{\linewidth}{@{\extracolsep{\fill}}c@{\hspace{2pt}\ablationsep\hspace{2pt}}
c c c@{}}
\toprule[0.8pt]
& \multicolumn{3}{c}{General repair instruction} \\
\cmidrule(l){2-4}
Model & \% Res. & \% Rec. & \% Cov. \\
\midrule[0.4pt]
GPT-6.1 Sol & 15.0\,(0.0) & 60.2\,(-6.3) & 37.2\,\textbf{(-32.2)} \\[2pt]
DS-V4.1-Flash & 10.0\,(-10.0) & 68.9\,(+4.9) & 32.7\,\textbf{(-37.1)} \\[2pt]
Qwen3.8-27B & 5.0\,(-5.0) & 56.5\,(-1.9) & 21.8\,\textbf{(-28.2)} \\[2pt]
\bottomrule[0.8pt]
\end{tabular*}
\endgroup
\caption{Effect of replacing cleanup instructions with a general repair request on \name{} Lite, with changes relative to the cleanup baseline in Table~\ref{tab:main-results}.}
\label{tab:ablation-cleanup-instruction}
\end{table}